\documentstyle[aps,preprint,epsf]{revtex}
\title {QCD Sum Rule and the Validity of Phenomenological Models}
\author{Tsuneki Matsuki
\footnote{email:matsuki@nt.phys.s.u-tokyo.ac.jp
				\hspace{5cm} FAX: (81)(3)56849642}
and Koichi Yazaki
\footnote{email:yazaki@phys.s.u-tokyo.ac.jp}}
\address{
Department of Physics,
University of Tokyo\\ Bunkyo--ku, Tokyo 113, Japan}
\date{1 October 1996}
\begin{document}                  
\begin{titlepage}
\maketitle
\begin{abstract}
The consistency of effective models with QCD is
investigated through the use of the QCD sum rule.
Taking the potential model for the heavy quark system,
we apply the method to two phenomenologically successful parameter sets,
and obtain the dependences of the model parameters 
on the QCD scale $\Lambda$.
Comparison with the expected scaling laws
allows us to reject one of the two sets.
The method is applicable to any model which reproduces 
the low lying spectra of hadronic systems.

\end{abstract}
\draft
\end{titlepage}
Heavy quark systems have been the most successful place 
in the study of strong interaction physics.
There exist many methods that give quantitative agreement with experiment,
both in the QCD based methods and in the QCD inspired models.
As for the former, 
the heavy quark effective theory has related various observables 
in the framework of perturbative QCD, 
and the QCD sum rule has been applied to various channels 
to calculate the masses of the heavy mesons.
As for the latter,
the non-relativistic quark model(QM) has proved itself
to be accurate enough to reproduce a vast number of physical observables,
with the bonus that it stays very close to intuition.
It is the mass of the heavy
quark itself that sets the scale in the meson masses, 
the role of the strong interaction being only to give the splitting
among them.

A problem about the QM is that there 
are usually several sets that seem to work equally well 
in a phenomenological sense.
Being a model, it is natural to expect that it reflects the
effective degrees of freedom when all the complicated degrees of freedom
in QCD are integrated out, and ultimately its parameters be calculated 
starting from the fundamental theory.
At this stage, no such calculation has been done,
which leaves us to rely either on their self-consistencies\cite{Eichten},
or on their phenomenological successes. 
In this paper, we propose the use of QCD sum rules 
to investigate the consistency of various models with QCD.
Evaluation of the vector current polarization function 
in the $c \bar c$ channel enables us
to exclude one of the two specific sets 
in the potential model for the charmonium system.
The method is in principle applicable for any phenomenological model,
the only requirement being that it reproduces the observed mass spectrum and
leptonic decay widths.

The QCD sum rule basically deals with the polarization function, 
evaluating it on the one hand by 
the operator product expansion(OPE) in perturbative QCD and by
a phenomenological model representing the observed spectrum
on the other.
Here we consider the vector polarization function 
\begin{equation}
 \Pi _{\mu \nu }(q^2)  \equiv  
  i\int_{-\infty }^{+\infty } {d^4x\;e^{iq. x}}\left\langle 0 \right|\, 
T\left(
{j_\mu \left( x \right)\, j_\nu \left( 0 \right)} \right)\, \left| 0 \right
\rangle 
=  ( q_{\mu} q_{\nu} - q^2g_{\mu \nu} ) \Pi(q^2)
\label{polarization}
\end{equation}
where $ j_{\mu}(x) = \bar {\psi}_c(x) \gamma_{\mu} \psi_c (x) $ and
 $q_{\mu}$ is the 4-momentum carried by the system.
Instead of extracting the physical masses by using the simplified 
spectral function,
we substitute a more realistic spectrum calculated by the model 
on the phenomenological side 
to see how the parameters in the model vary with the
quantities appearing in the OPE. 
More specifically, 
we calculate the variation of $\Pi(Q^2)$ with the QCD scale parameter 
$\Lambda$
\footnote{$\Lambda$ stands for $\Lambda_{\mbox{mom}}$ 
in this particular case.},
and fit it for a certain range of $Q^2$
with the variations in the QM side with respect
to the parameters of the model.
It is essential that we fit the $change$ in the polarization function 
and not the function itself, 
since the latter is usually 
saturated by the first few resonances, and is insensitive to the parameters
as long as they accurately 
reproduce the first few spectra.
The validity of the model parameters is investigated through their 
$\Lambda$ dependences.	

For our analysis, we take the simple 'Cornell' type potential
\begin{equation}
V(r)= -{a \over r}+\kappa \ r+V_0,\;
\label{potential}
\end{equation}
with two sets of parameters as our candidates.
Set A has four adjustable parameters and 
reproduces $m_{\psi(1S)}$,$m_{\psi(2S)}$,
$\Gamma_{\psi(1S) \rightarrow e^+e^- }$
and $\Gamma_{\psi(2S)\rightarrow e^+e^- }$,
with the radiative correction factor $(1-{4 \over \pi} a)$ 
for the widths
The parameters take the values
$$\{m=2.04[\mbox{GeV}],a=0.579[1],\kappa=0.172[\mbox{GeV}^2],
V_0=-1.12[\mbox{GeV}]\} ,$$
which are essentially the ones employed by \cite{Eichten}.
Set B, on the other hand, has one restriction $a={4\over3} \alpha_s=0.27$, 
and is thus adjusted to reproduce only
$m_{1S}$,$m_{2S}$ and $\Gamma_{\psi(1S) \rightarrow e^+e^- }$.
The parameters are given by $$\{m,a,\kappa,V_0\}=\{1.78,0.270,0.222,-1.00\}$$
\cite{Barbieri}.
Their characteristics are summarised in Table \ref{model}.

The former is naturally better in reproducing the experimental data, and is 
the most frequently used set in modern quark model analyses.
The latter on the other hand has the strong point that 
the Coulomb force is directly correlated with the
one-gluon exchange of QCD.
Set B with the coupling with $D \bar D$ channel will be discussed later.
With these models at hand, we calculate
\begin{equation}
\Pi(Q^2) 
={1 \over \pi }\int\limits_0^\infty
{ds{{\mbox{Im}\Pi (s)} \over {(s+Q^2)}}} ,
\label{mom}
\end{equation}
with Im$\Pi(s)$ given by
$\mbox{Im}\Pi (s)=\sum\limits_n
{{\pi  \over {3s}}\left| {\left\langle 0
\right|j^\mu (0)\left| {0, n } \right\rangle } \right|^2
\delta (s-M_n^2)} $.
$\left| k,n \right\rangle$ denotes the $n$th bound state with momentum $k$,
with the normalization given by
$\left\langle {{p, n }} \mathrel{\left | {\vphantom {{p, n }
 {p', n' }}} \right.  \kern-\nulldelimiterspace} {{p', n' }}
\right\rangle
=(2\pi )^32p_0\delta ^3(p-p')\delta_{n n'} $. 
$n$ includes the possible polarizations of the state.
The matrix elements are evaluated with the use of the 
van Royen-Weisskopf formula\cite{Royen}.

On the other hand, the OPE side of the QCD sum rule is calculated in the
usual manner \cite{Reinders},
$$ \Pi(Q^2) = C_0 (Q^2) +  C_G(Q^2)  
\left\langle 0 \right|{\alpha_s \over \pi}
G_{\mu \nu }G_{\mu \nu }\left| 0 \right\rangle + ....$$
where the dots indicate higher dimensional contributions
\footnote{We take into account upto operators of dimension 8.
Heavy quark condensates are reduced to gluon condensates 
with the use of the heavy quark mass expansion\cite{Reinders,quark}.
}.
Neglecting light quarks, we have two parameters on this side, i.e.
the QCD scale parameter $\Lambda$ and the heavy quark current mass, $m_c$.
We are here interested in the variants of $\Pi$ with respect to
$\Lambda$, with $m_c$ fixed.
The essential point is that we control the strength of
the interaction solely through the parameter $\Lambda$.
This parameter comes in through the quantities $\alpha_s$ and 
the gluon condensates 
( $\left\langle 0 \right|{\alpha_s \over \pi}
G_{\mu \nu }G_{\mu \nu }\left| 0 \right\rangle$ etc.).
$\left\langle 0 \right|{\alpha_s \over \pi}
G_{\mu \nu }G_{\mu \nu }\left| 0 \right\rangle$		
behaves as $\Lambda^4$ since it has no anomalous dimension 
(we are neglecting light quarks) and $\alpha_s$ will behave as
$1/\mbox{ln}(m_c/\Lambda)$. 
It is easy to see that taking $\Lambda \rightarrow 0$ will give 
0 for both quantities, which means that the interaction is turned off
and we are left with only the free particle contribution to the
polarization function.

Now we can compare both sides concerning their variants with respect to 
$\Lambda$,
\begin{equation}
{{\partial \Pi (Q^2)} \over {\partial \Lambda }}\left| {_{m_c}} \right.
={{\partial \Pi (Q^2)} \over {\partial m}}\left| {_{a,\kappa }} \right.
{{\partial m} \over {\partial \Lambda }}\left| {_{m_c}} \right.+{{\partial \Pi
(Q^2)} \over {\partial a}}\left| {_{m,\kappa }} \right.{{\partial a} \over
{\partial \Lambda }}\left| {_{m_c}} \right.+{{\partial \Pi (Q^2)} \over
{\partial \kappa }}\left| {_{a,m}} \right.{{\partial \kappa } \over {\partial
\Lambda }}\left| {_{m_c}} \right..
\end{equation}
The derivatives of $\Pi(Q^2)$ with respect to the parameters are
calculated by numerical differentiation,
at some fixed values of $\Lambda,m_c,a$, etc.
The coefficients 
( ${{\partial m} \over {\partial \Lambda }}\left| {_{m_c}}\right.$,... )
are obtained as a result of the fit in the region 
$Q^2=0.5 \sim 3.0 \mbox{GeV}^2$ \cite{Reinders}.
In the actual calculation, we evaluate the moments of $\Pi(Q^2)$, i.e. 
$\Pi_n(Q^2) \equiv {1 \over n!}(-{d \over {dQ^2}})^n\Pi (Q^2)$
since this has an effect of emphasizing the low energy part of
the physical spectrum\cite{Shifman}.
We also evaluate for the parameter $O$,
\begin{equation}
d_O = {{\partial ( \mbox{ln} O ) } \over {\partial (\mbox{ln} \Lambda ) }} 
\end{equation}
instead of 
${{\partial O} \over {\partial \Lambda }} $
for reasons of convenience.
This implies that the parameter behaves as $O\sim\Lambda^{d_O}$
near the actual value of $\Lambda$.

The fit is obtained in the simplest way through a linear 
$\chi^2$ fit formula (matrix inversion).
We have found that the matrix is practically singular, 
which means that the effect of the parameters are redundant 
when reflected in the behavior of the polarization function.
This is not surprising when we recall that
the function itself is saturated by the first few resonances,
which leaves us with only $2n$ degrees of freedom where $n$ is 
the number of resonances required to saturate the function.
Actually, the lowest one is dominant and
we found that only two out of four parameters were independent.
This forces us to fix two of the $\Lambda$ dependences of the parameters,
which we choose to be $\kappa \sim \Lambda^2$ and $V_0 \sim \Lambda$.
This corresponds to taking the quenched approximation for the
interaction, which is in agreement with 
the spirit of the quark model
\footnote{Of course we do not explicitly handle
the dynamical (radiative) effects of the gluons\cite{Voloshin}.
We take the view that their effects are integrated out and 
arise as a change in the potential and the quark mass when we
restrict ourselves to the two fermion sector.}.

\begin{figure}
\epsfbox{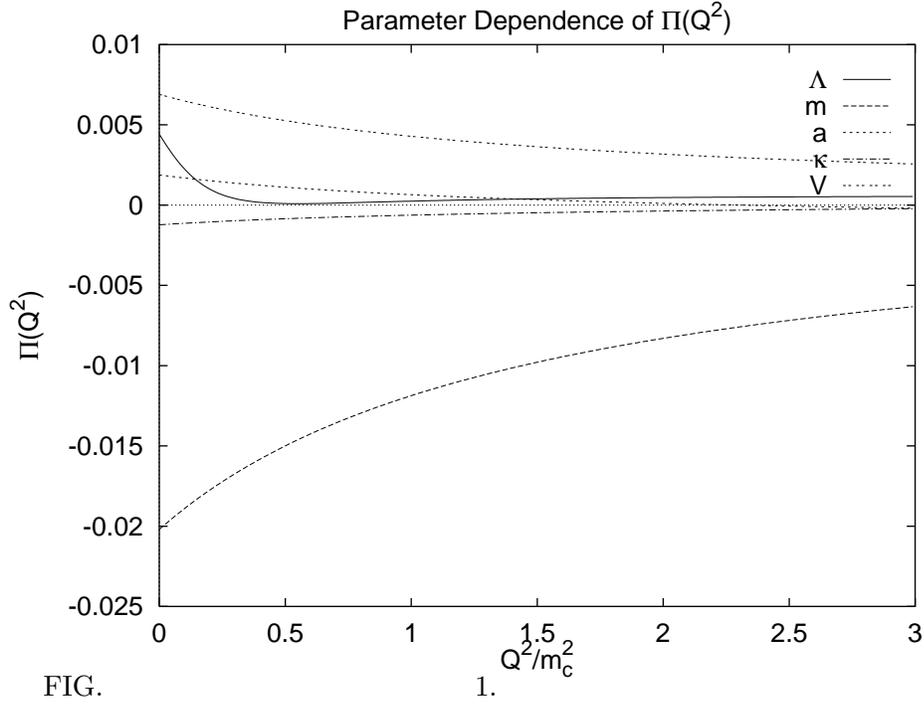}
\caption{
Variation with respect to $\Lambda,m,a,\kappa$ and $V_0$.
Plotted are ${\partial \mbox{ln}\Pi_{OPE}}/{\partial \mbox{ln}\Lambda},
{\partial \mbox{ln}\Pi_{QM}}/{\partial \mbox{ln}m}$ 
and its counterparts.
}
\label{naive}
\end{figure}

For set A,
fitting the curves shown in figure \ref{naive} gives 
the scaling behaviors of the model parameters expressed as
$O \sim \Lambda^{d_O}$, with $d_O$
$$\{d_m,d_a,d_{\kappa},d_v\}=\{0.28 \pm 0.02,0.6 \pm 0.3,2.,1.\}$$
where $d_{\kappa}$ and $d_v$ are inputs.
The error is due to the following ambiguities \cite{Reinders};
\begin{itemize}
\item{Region of $Q^2$ where the fit is performed.}
\item{Which moment we take. : $\Pi_6 \sim \Pi_{10}$}
\item{Value of $\alpha_s(4m_c^2)$, or equivalently, 
$\Lambda_{\mbox{mom}}:130 \sim 220$[MeV] }
\item{Value of $\left\langle 0 \right|{\alpha_s \over \pi}
G_{\mu \nu }G_{\mu \nu }\left| 0 \right\rangle :
(360 \pm 50[\mbox{MeV}])^4$	}
\end{itemize}
Although we know of no correct formula for the analytic behavior of 
$m$ against $\Lambda$,
we can give an estimate for the $\Lambda$ dependence
from the following arguments.
First, it is natural to expect that $m$ reduces to $m_c$ 
(the 'on shell' mass \cite{charmmass} ) 
in the limit $\Lambda \rightarrow 0$, which is the free limit.
This determines the sign of $d_m$.
Second, from dimensonal analysis we expect that
$\Delta m = m - m_c \sim c_m \Lambda$ 
where $c_m$ is some constant of proportionality. 
This requires that 
\begin{equation}
d_m = {{\partial (\mbox{ln} m) } \over {\partial (\mbox{ln} \Lambda) }} = 
{{\Delta m} \over m}
\end{equation}
which, after substituiton of $ m = 2.$[GeV] and $m_c = 1.5$[GeV], gives
$d_m = 0.25$ which is clearly close to the value of set A.
(allowing a dependence of $m \sim \mbox{ln}( \Lambda/ m_c )$ does
 not alter the essence of this estimate.)
Also the dependence of $a$ on $\Lambda$ is consistent with 
that of the QCD $\alpha_s$ parameter($d_a \sim d_{\alpha_s}$).

On the other hand, for set B we obtain
$$\{d_m,d_a,d_k,d_v\}=\{-0.2 \pm 0.1,-4.9 \pm 1.9,2.,1.\}$$
We notice that they fulfil neither of the requirements discussed above.
They are not appropriate firstly since their signs are the opposite,
and secondly because we would not expect such a strong dependence of
$a$ on $\Lambda$.

\begin{figure}
\epsfbox{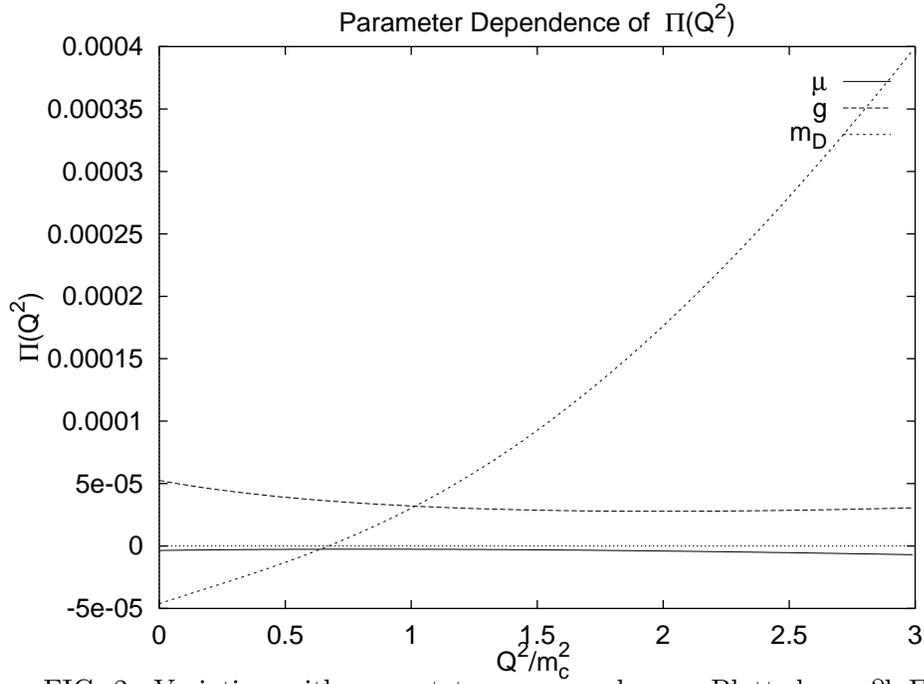}
\caption{Variation with respect to $\kappa,\mu,g$ and $m_D$.
Plotted are 
${\partial \mbox{ln}\Pi_{QM}}/{\partial \mbox{ln}\kappa},
{\partial \mbox{ln}\Pi_{QM}}/{\partial \mbox{ln}\mu}$
and its counterparts.}
\label{DD}
\end{figure}

These results clearly show that we should take set A as our phenomenological
model in order to meet the requirements of QCD.
This is rather contrary to our intuition since it seems that 
set B is more easily justified from the viewpoint of perturbative QCD.
We have considered the following facts that might be
the reasons for the failure of set B:
\begin{enumerate}
\item{
Set B does not reproduce $\Gamma_{\psi(2S) \rightarrow e^+e^-}$
(Table \ref{model}). 
The effect of the second bound state accounts for a non-negligible 
portion of the polarization function.}
\item{
The 'radiative' correction $(1-{4 \over \pi}a)$ differs largely 
in the two cases.
Taking $a$ to be the value of set A, this gives a correction
of more than $50\%$ of the magnitude.
While this is often considered as a weak point in a quark model analysis,
we on the other hand have obtained a result that supports this factor.}
\end{enumerate}
To investigate the first effect,
we have performed a coupled channel analysis
similar to the one in \cite{Eichten}.
For simplicity, we have taken only the $D \bar D$ channel into account,
with the coupling potential 
$$\left \langle { c\bar c \left| V \right| D \bar D} \right\rangle
= g\ e^{-\mu r}{{r_i} \over r}\varepsilon _i,$$
where $\varepsilon _i$ denotes the polarization vector of the 
$c\bar c$ system.
This allows us to reproduce the decay width 
$\Gamma_{\psi(2S) \rightarrow e^+e^-}$.
Going through the same analysis but now with a larger parameter space
$\{m,a,\kappa,V_0,\mu,g,m_D\}$, we obtain figure \ref{DD},
which shows the contribution from 
the changes in the model parameters
$\mu,g,m_D$ and $\kappa$ respectively. 
The difference in their magnitudes is clear.
Requiring $\mu,g$ and $m_D$ to take appropriate values 
($d_{\mu}=d_{g}=1,d_{m_D}=.2$)
\footnote{$d_{m_D}$ is evaluated as $\{m_D-(m_c+m_u)\}/m_D=0.2$.}
, they clearly have only a small effect on the fit.
We gain the values 
$$\{d_m,d_a,d_{\kappa},d_v\}=\{-0.15 \pm 0.06,-3.8 \pm 0.7,2.,1.\},$$
which is essentially unchanged from the previous fit.
Thus, the modified set B(including the coupled channel effect) works 
equally well phenomenologically as the set A but the scaling behavior
of the parameters is still inconsistent with that expected from QCD.

One can easily diminish the effect of the factor 
$(1-{4 \over \pi} a)$ 
by simply replacing $\Pi_8$ with the ratio $\Pi_{n+1}/\Pi_n$. 
The previous result obviously satisfies the fit, 
but due to the cancellation there might still
be a set of parameters that is consistent with OPE.
Notice that we have even fewer degrees of freedom (namely one),
since taking a ratio means losing the information on the magnitude.
The analysis gives the result 
$$\{d_m,d_a,d_{\kappa},d_v\}=\{0.10 \pm .04,0.2,2.,1.\},$$
where we have substituted $d_a=0.2$.
Varying $d_a$ gives a change in the result which is included in the error.
We have thus obtained 
the scaling behavior of set B close to that of set A,
but this is only due to the fact that we are looking at a
fit with fewer requirements. Requiring the individual moments
to behave correctly will 
force the parameters to go to the inappropriate value.
This indicates that it is indeed the 'radiative correction' factor 
$(1- {4\over \pi}a)$ that prohibits 
the use of set B.
In other words, the factor is essential in achieving the consistency with OPE.

In this paper, we have shown that it is set A of the model parameters
that is consistent with the OPE, and is therefore appropriate for use
in calculating physical quantities.
This is in agreement with the choice of \cite{Eichten},
who justifies this set through comparison with experimental data. 
We have also shown that it is the 'radiative correction' 
factor itself that was responsible for this result, 
although its 'physical' meaning remains unclear.
Our analysis shows that in one way or the other, the total magnitude
of Im$\Pi(s)$ must be appreciably modified from the simple potential 
picture.

The method is applicable in exactly the same fashion for a model
that reproduces the spectrum and the leptonic decay widths.
Applying finite temperature QCD sum rules,
the method can also be used to extract the temperature dependences of 
the model parameters.
In this case, one has to substitute two of the temperature dependences.

T.M. wishes to thank N. Ishii for various ideas and discussions,
S. Takeuchi for continuous encouragement, and finally 
K.Yamashita and T.Mizusaki for tips on numerical computation.
    This work was supported in part by Grant-in-Aid for General Scientific
    Research (No.  04804012) and by Grant-in-Aid for Scientific Research on
    Priority Areas (No.  05243102) from the Ministry of Education,  Science
    and Culture.

\clearpage

\begin{table}
\caption{Calculated Spectrum}
\label{model}
\begin{tabular}{lcccc}
\tableline
& set A & set B & B with $D\bar D$ & Exp. \\
\tableline
$m_{1S}$ [GeV] & 3.097 & 3.097 & 3.097 & 3.097 \\
$m_{2S}$ [GeV] & 3.685 & 3.685 & 3.685 & 3.685 \\
$m_{1D}$ [GeV] & 3.82  & 3.775 & 3.75  & 3.770 \\
$m_{4040}$[GeV]& 4.09  & 4.14  & 4.2   & 4.040 \\
$\Gamma_{\psi(1S) \rightarrow e^+e^- }$[keV] & 5.26 & 5.26 & 5.26 & 5.26 \\
$\Gamma_{\psi(2S) \rightarrow e^+e^- }$[keV] & 2.14 & 3.0 & 2.14 & 2.14 \\
\tableline 
\end{tabular}
\end{table}

\clearpage

\pagestyle{empty}

\end{document}